\documentclass[referee]{raa_rb}

\usepackage{graphicx,times,multirow}
\usepackage{natbib}

\usepackage{amssymb,amsmath}
\bibpunct{(}{)}{;}{a}{}{,}

\newcommand{\eventa}{KMT-2025-BLG-1314}
\newcommand{\eventb}{KMT-2025-BLG-1392}
\newcommand{\thetae}{\theta_{\rm E}}

\usepackage[pagebackref=true]{hyperref}

\begin{document}

   \title{KMT-2025-BLG-1314 and KMT-2025-BLG-1392: two microlensing planetary/brown-dwarf candidates analyzed with differentiable code}

 \volnopage{ {\bf 20XX} Vol.\ {\bf X} No. {\bf XX}, 000--000}
   \setcounter{page}{1}

   \author{  %
   Haibin Ren \inst{1}, Weicheng Zang\inst{2}, Wei Zhu \inst{1}, Yoon-Hyun Ryu\inst{3}, Yuchen Tang\inst{2}, Jiyuan Zhang\inst{1}, Michael D. Albrow\inst{4}, Sun-Ju Chung\inst{3}, Andrew Gould\inst{5,6}, Cheongho Han\inst{7}, Kyu-Ha Hwang\inst{3}, Youn Kil Jung\inst{3,8}, In-Gu Shin\inst{2}, Yossi Shvartzvald\inst{9}, Hongjing Yang\inst{10,2}, Jennifer C. Yee\inst{11}, Dong-Jin Kim\inst{3}, Chung-Uk Lee\inst{3}, Byeong-Gon Park\inst{3}, Yunyi Tang\inst{1}, Dan Maoz\inst{12}, Shude Mao\inst{2}, Qiyue Qian\inst{2,1}
   }

\institute{Department of Astronomy, Tsinghua University, Beijing 100084,
China;\\
\and
Department of Astronomy, Westlake University, Hangzhou 310030, Zhejiang Province, China\\
\and
Korea Astronomy and Space Science Institute, Daejon 34055, Republic of Korea\\
\and
University of Canterbury, School of Physical and Chemical Sciences, Private Bag 4800, Christchurch 8020, New Zealand\\
\and
Max-Planck-Institute for Astronomy, K\"onigstuhl 17, 69117 Heidelberg, Germany\\
\and
Department of Astronomy, Ohio State University, 140 W. 18th Ave., Columbus, OH 43210, USA\\
\and
Department of Physics, Chungbuk National University, Cheongju 28644, Republic of Korea\\
\and
National University of Science and Technology (UST), Daejeon 34113, Republic of Korea\\
\and
Department of Particle Physics and Astrophysics, Weizmann Institute of Science, Rehovot 7610001, Israel\\
\and
Westlake Institute for Advanced Study, Hangzhou 310030, Zhejiang Province, China\\
\and
Center for Astrophysics $|$ Harvard \& Smithsonian, 60 Garden St., Cambridge, MA 02138, USA\\
\and
School of Physics and Astronomy, Tel-Aviv University, Tel-Aviv 6997801, Israel\\
\vs \no
   {\small Received 20XX Month Day; accepted 20XX Month Day}
}

\abstract{
Analysis of binary-lens microlensing events typically requires intensive computation because of the multimodal and complex posterior distributions. With the recent development of the JAX-based differentiable binary-lensing modeling package \texttt{microlux}, we present an analysis of two microlensing events with planet/brown-dwarf candidates, KMT-2025-BLG-1314 and KMT-2025-BLG-1392. Both events exhibit the ``Close/Wide'' degeneracy, and KMT-2025-BLG-1314 suffers from the ``Planet/Binary'' degeneracy and a recently recognized ``Point/Finite'' degeneracy among the planetary solutions. For KMT-2025-BLG-1314, the binary mass ratio is \(\log q \sim -3.5\) for the planetary solutions and \(\log q > -1.5\) for the binary solutions, while for KMT-2025-BLG-1392 we find \(\log q \sim -1.3\). We show that for the analysis of KMT-2025-BLG-1314, Hamiltonian Monte Carlo (HMC), enabled by \texttt{microlux}, provides robust parameter inference and outperforms traditional Markov chain Monte Carlo (MCMC) methods in the presence of bimodal posteriors.
}

   \authorrunning{ }            %
   \titlerunning{ }  %
   \maketitle

\section{Introduction}           %
\label{sect:intro}

Gravitational microlensing has been widely applied to a broad range of astrophysical problems, most notably the detection of exoplanets and stellar binaries \citep{mao1991,Andy1992}. As of 2026 January \footnote{Based on a query made in January 2026}, more than 263 exoplanets have been discovered via microlensing, according to the NASA Exoplanet Archive \citep{Christiansen2025}. This growing planet sample enables statistical studies of exoplanet demographics \citep{zhu2021}, especially the cold planet population (e.g., \citealt{Wise,Zang2025}).

Almost all of the microlensing planets are discovered by ground-based surveys toward the Galactic bulge. These include the Optical Gravitational Lensing Experiment (OGLE, \citealt{OGLEIV}), the Microlensing Observations in Astrophysics survey (MOA, \citealt{Sako2008}), and the Korea Microlensing Telescope Network (KMTNet, \citealt{kmtnet2016}). Upcoming space-based missions, including the Nancy Grace Roman Space Telescope \citep{penny2019}, the Earth 2.0 Microlensing Telescope \citep{CMST,Ge2022}, and the Chinese Space Station Telescope (CSST, \citealt{yan2022}), are expected to increase the number of microlensing planetary detections to the thousands and systematically discover currently rare systems/objects, including dark lenses \citep{sahu2022,lam2022,mroz2022}, free-floating planets \citep{Mroz2017a,Gould2022_FFP_EinsteinDesert,Sumi2023}, and lenses with multiple components (e.g., \citealt{gaudi2008}).

Along with the success in the discovery front, the detailed analysis of binary-lens and multiple-lens microlensing events remains a challenge. The inherently complex parameter space is characterized by strong correlation/continuous degeneracy (e.g., \citealt{gaudi&gould1997}), multi-modality (e.g., the well-known close-wide degeneracy, \citealt{Griest1998, Dominik1999, an2005}), and high-order effects (e.g., the microlensing parallax and xallarap effects, \citealt{Gould1992,griest&hu1992}). Typical modeling approaches combine brute-force searches, template matching \citep{Liebig2015,bozza2024}, and other global optimization techniques to explore the multi-modal parameter space, followed by Markov chain Monte Carlo (MCMC) sampling to characterize the local posterior distributions and estimate parameter uncertainties. However, conventional MCMC methods often struggle with posterior distributions that exhibit strong non-linear correlations, high dimensionality, and multiple modes, leading to inefficient sampling, strong autocorrelations, and poor convergence across different chains, ultimately degrading the accuracy of the inferred parameters.

The development of the automatic differentiation technique has made it possible to obtain accurate gradients even for complicated models. This then promotes the use of gradient-based sampling techniques like Hamiltonian Monte Carlo (HMC;  \citealt{duaneHybridMonteCarlo1987, neal2011, Hoffman2011}) in many modern probabilistic programming languages, such as \texttt{Stan} \citep{carpenter2017}, \texttt{Numpyro} \citep{phan2019composable}, and \texttt{Blackjax} \citep{cabezas2024blackjax}). An emerging trend in astronomy is to develop differentiable code based on automatic differentiation packages (e.g., \texttt{Julia} \citealt{Julia-2017} and \texttt{Jax} \citealt{jax2018github}), and successful applications are seen across a wide range of research fields (an incomplete list can be found here \footnote{\url{https://juliaastro.org/home/}} \footnote{\url{https://github.com/JAXtronomy/awesome-JAXtronomy}}). In microlensing, several differentiable models have been developed, including \texttt{caustics} \citep{bartolic2023a}, \texttt{microlux} \citep{ren2025}, and \texttt{microJAX} \citep{Miyazaki2025}. Among these, \texttt{microlux} is unique for its specially designed error estimator with adaptive contour integration \citep{bozza2010, bozza2018,bozza2025}. This can ensure the convergence of both the microlensing magnification and its derivatives. So far, applications of these differentiable models are limited to either synthetic data or well-studied historical events. %

Here we present the analysis of two microlensing binary-lens events, KMT-2025-BLG-1314 and KMT-2025-BLG-1392, based on the differentiable model of \texttt{microlux}. Both events were discovered by KMTNet and followed up by the Microlensing Astronomy Probe (MAP) program using the Las Cumbres Observatory Global Telescope (LCOGT; \citealt{Brown2013}) network. The MAP program targets high-magnification microlensing events and has observed about a dozen planets to date, including two systems that each contains two planet-like companions \citep{Zang2021,KB221818}.

This paper is organized as follows: Section~\ref{sec: observation} describes the survey and follow-up observations, Section~\ref{sec: light_curve_analysis} presents the binary-lens single-source (2L1S) and single-lens binary-source (1L2S) modeling, Section~\ref{sec:source_lens_properties} shows the source and lens properties with the color-magnitude diagram (CMD) and Bayesian analysis, respectively. Finally, in Section~\ref{sec:disscussion}, we discuss the lessons learned from the differentiable modeling, focusing especially on the HMC method. 

\section{Observations} \label{sec: observation}

The microlensing events \eventa\ and \eventb\ were first alerted by the KMTNet AlertFinder system \citep{kim2018b} on 2025 June 10 and June 17, respectively. The observations were obtained with 1.6~m telescopes equipped with $4~\mathrm{deg}^2$ mosaic CCD cameras at Cerro Tololo Inter-American Observatory (CTIO) in Chile (KMTC), the South African Astronomical Observatory (SAAO) in South Africa (KMTS), and Siding Spring Observatory (SSO) in Australia (KMTA). Most KMTNet images were taken in the $I$ band, with $\sim 9\%$ obtained in the $V$ band to measure the source color.

The source star of \eventa\ is located at equatorial coordinates $(\alpha,\delta) = (17{:}30{:}12.47,\,-30{:}31{:}14.81)$, corresponding to Galactic coordinates $(\ell,b) = (-3.1299^\circ,\,+1.9777^\circ)$. This event lies in the KMTNet BLG11 field, which is monitored with an observing cadence of $0.4~\mathrm{hr}^{-1}$ for KMTC and $0.3~\mathrm{hr}^{-1}$ for KMTA and KMTS. The source star of \eventb\ is located at equatorial coordinates $(\alpha,\delta) = (17{:}43{:}28.63,\,-26{:}29{:}26.41)$, corresponding to Galactic coordinates $(\ell,b) = (+1.8371^\circ,\,+1.6827^\circ)$. This event lies in the KMTNet BLG18 field, with an observing cadence of $1.0~\mathrm{hr}^{-1}$ for KMTC and $0.8~\mathrm{hr}^{-1}$ for KMTA and KMTS. The KMTNet field layout is shown in Figure~12 of \citet{KMTeventfinder}.

Following alerts issued by the KMTNet HighMagFinder system \citep{KB210171}, follow-up observations were conducted by LCOGT and KMTNet. HighMagFinder is designed to identify high-magnification KMTNet events using real-time photometry before they reach a magnification threshold of $A_{\rm thresh} = 25$, enabling timely follow-up observations. For \eventa, LCOGT observations were carried out using the 1\,m telescopes at SSO and SAAO, which we designate LCOA and LCOS, respectively. For \eventb, MAP used the two 1\,m telescopes at LCOGT-SSO (LCOA01 and LCOA02) and the 1\,m telescope at the LCOGT Teide Observatory in Tenerife, Spain (LCOT), to monitor the event during its peak. All LCOGT observations were obtained in the $I$ band. KMTNet employed its ``auto-followup'' mode to increase the observational cadence near the peak. For \eventa, the KMTA cadence was raised from $0.3~\mathrm{hr}^{-1}$ to $\Gamma = 3.3~\mathrm{hr}^{-1}$ through the substitution of BLG43 observations with those from BLG11. A similar strategy was applied to KMTS, for which replacing BLG03 with BLG11 observations resulted in the same increase in cadence, from $0.3~\mathrm{hr}^{-1}$ to $\Gamma = 3.3~\mathrm{hr}^{-1}$. For \eventb, the KMTA observing cadence increased from $0.8~\mathrm{hr}^{-1}$ to $\Gamma = 3.8~\mathrm{hr}^{-1}$ following the replacement of BLG43 observations with BLG18 observations.

The KMTNet and LCOGT data were both reduced using the pySIS pipeline \citep{pysis,Yang_TLC,Yang_TLC2}, which is based on the difference image analysis (DIA) technique \citep{Tomaney1996,Alard1998}. The data error bars were rescaled following the method in \citet{yee2012}.

\section{Light-curve analysis}\label{sec: light_curve_analysis}

\subsection{Preamble}\label{sec:preamble}
We start from the standard 2L1S model, which is described with seven parameters. These include three point-source point-lens (PSPL, \citealt{paczynski1986}) parameters: \(t_0\), the time of closest lens-source alignment; \(u_0\) the closest distance of the source to the lens center of mass in the unit of angular Einstein radius \( \theta_\mathrm{E} \); \( t_\mathrm{E} \), the Einstein radius crossing time. This timescale is related to the Einstein radius \( \theta_\mathrm{E}\) and the total lens mass \(M_\mathrm{L}\) by:
\begin{equation}
    t_E \equiv \frac{\theta_\mathrm{E}}{\mu_\mathrm{rel}}; \qquad
    \theta_\mathrm{E} \equiv \sqrt{\kappa M_\mathrm{L} \pi_\mathrm{rel}}.
\end{equation}
Here \( \kappa \equiv \frac{4G}{c^2 \mathrm{au}} \simeq 8.144\mathrm{mas}/M_\odot \), and \( (\pi_\mathrm{rel}, \mu_\mathrm{rel}) \) are the lens-source relative parallax and proper motion \citep{Gould2000}. Another three parameters are required to describe the binary lens geometry: $s$, the projected planet-to-star separation in units of $\thetae$; $q$, the planet-to-star mass ratio; and $\alpha$, the angle between source trajectory and binary lens axis. The last parameter \(\rho \) is the angular size of the source, \(\theta_\star\), in units of $\thetae$. The blending effect is described by two linear parameters for each data set $i$: the source flux $f_{{\rm S},i}$ and the blending flux $f_{{\rm B},i}$. The observed flux for each data set is then given by 
\begin{equation}
    F_i(t) = f_{{\rm S},i} A(t) + f_{{\rm B},i} .
\end{equation}
For the light curve in each observational site, we first use a binary lens model to calculate the magnification and then use linear regression to determine the blending parameters. 

We start the modeling process with a parameter search in a dense grid in the $(s,q,\alpha )$ space. At each grid point, the initial guesses for the parameters $(s, q, \alpha)$ are given by the values of the grid point, and values for the other parameters \((t_0,u_0,t_\mathrm{E},\rho)\) are taken from the PSPL fit. The model $\chi^2$ is minimized via the Covariance Matrix Adaptation Evolution Strategy (CMA-ES) algorithm \footnote{\url{https://github.com/CMA-ES/pycma}} from \texttt{pycma} \citep{hansen2019pycma}. During the minimization, values of the parameters \((s,q) \) are fixed, whereas the other parameters \((t_0,u_0,t_\mathrm{E},\rho,\alpha)\) are allowed to vary. The best solutions in each grid are then polished with the Nelder-Mead simplex algorithm \citep{nelderSimplexMethodFunction1965, gaoImplementingNelderMeadSimplex2012} with all parameters as free. Because no model gradients are used in either optimization algorithm, we use the \texttt{VBMicrolensing}\citep{bozza2010,bozza2018,bozza2025} to compute the 2L1S magnification at this stage.

For high-magnification events, the limb-darkening effect is important for estimating the source radius. We use the linear limb-darkening law described by  
\begin{equation}
    S_\lambda(\nu) = S_\lambda(0)[1-a_\lambda(1-\nu)] \quad \nu = \sqrt{1-\frac{r^2}{\rho^2}},
\end{equation}
where the \(S_\lambda(0)\) is the surface brightness at the center of the source, \(r\) and \(\rho\) are the distance to the center of the source and the source radius, respectively, and \(a_\lambda\) is the limb-darkening coefficient at wavelength \(\lambda\)
We infer the effective temperature $T_{\rm eff}$ in Section~\ref{sec:source properties} and obtain the limb-darkening coefficients based on the stellar model in \citet{Claret2011}. 

After the locations of the probable solutions are identified from the grid search, we are then set to sample the posterior distributions of the model parameters. The 2L1S posteriors for these two high-magnification events are obtained using HMC, which is a gradient-based variant of MCMC. Compared to the traditional MCMC, HMC has several advantages, including the ability to efficiently draw samples with weaker autocorrelation while maintaining a high acceptance rate, even in high dimensions \citep{neal2011}. We use the \texttt{microlux} code for the 2L1S model at this stage. In the recent version 0.2.0, \texttt{microlux} adopts the new polynomial coefficient calculation of \citet{Wang2025} to boost the robustness for planetary and extreme binary cases, which will be the situation we face for the two events under consideration.

To maximize HMC performance, we utilize the local covariance obtained from the Fisher information matrix to reparameterize the binary microlensing parameters, thereby mitigating the strong correlation between certain parameters. This method has been applied in historical events modeling \cite{ren2025, Miyazaki2025}. Here, we follow a similar scheme, which nevertheless differs in some aspects, to implement the reparameterization. Specifically, we initially draw unconstrained and uniform samples from the latent space $\boldsymbol{ \beta'}$, implemented by \texttt{ImproperUniform} in \texttt{Numpyro}.  The uniform and unconstrained features are chosen to avoid the prior information introduced in the later reparameterization. The true parameters are given by a triangular affine transformation 
\begin{equation}
    \boldsymbol{\beta} = \boldsymbol{\mu}+\boldsymbol{L} \boldsymbol{\beta'}, 
\end{equation}
where $\boldsymbol{\mu}$ is the initial guess, $\boldsymbol{L}$ is a lower triangular matrix from Cholesky factorization of the covariance matrix $\boldsymbol{LL}^T = \boldsymbol{\Sigma}$. To constrain the transformed parameters within physical bounds without disrupting the continuous gradients required by HMC, we impose a soft-boundary penalty in likelihood via the \texttt{softplus} function. This reparameterization effectively preconditions the sampling space, equivalent to adapting the HMC mass matrix (metric) to the local covariance, thereby accelerating the warm-up phase.

For both events, the bump-type anomaly could also be caused by a 1L2S model, which can mimic the 2L1S model \citep{gaudi1998}. Thus, we also try to fit a 1L2S model to the data. A 1L2S model is defined with two sets of PSPL parameters, but with a common timescale, \(t_\mathrm{E}\), which correspond to two background source stars. The 1L2S total magnification at waveband \( \lambda \), \(A_\lambda (t)\), is given by \citep{hwang2013}
\begin{equation}\label{equ: 1L2S}
    A_{\lambda}(t) = \frac{A_{1}(t)f_{{\rm S}, 1,\lambda} + A_{2}(t)f_{{\rm S}, 2,\lambda}}{f_{{\rm S}, 1,\lambda} + f_{{\rm S}, 2,\lambda}} = \frac{A_{1}(t) + q_{f,\lambda}A_{2}(t)}{1 + q_{f,\lambda}}, 
\end{equation}
where
\begin{equation}
    q_{f,\lambda} \equiv \frac{f_{{\rm S}, 2,\lambda}}{f_{{\rm S}, 1,\lambda}}
\end{equation}
is the ratio of the two source fluxes at wavelength \( \lambda\). The quantity \(A_j(t)\) represents the magnification given by the $j$-th source. We use \texttt{VBMicrolensing} to calculate the 1L2S model with the finite source effect.

\subsection{KMT-2025-BLG-1314}\label{sec: 1314}

\begin{figure}
   \centering
   \includegraphics[width=\linewidth]{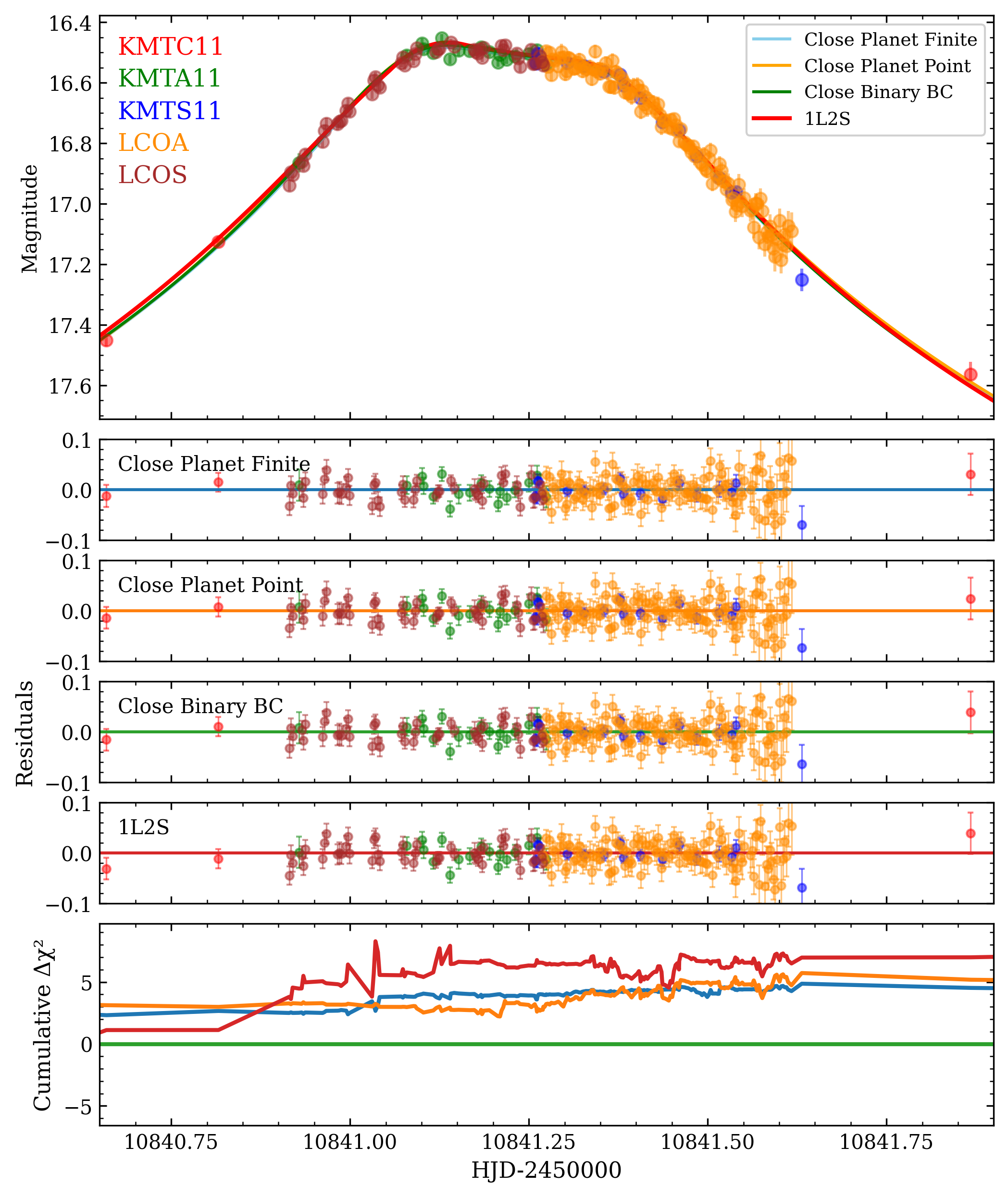}
   \caption{Light curve and the best-fit model of the microlensing event \eventa. Different data sets are shown in different colors. The ``Close Binary BC'' solution has the lowest $\chi^2$. The bottom panel shows the residual distribution for four different solutions relative to the ``Close Binary BC'' solution. Different data sets are aligned to the KMTC11 $I$-band data.}
   \label{fig:1314}
\end{figure}

Figure \ref{fig:1314} displays the observed data of \eventa, which has an asymmetric feature or a double-bump feature over the peak of a PSPL model.

We include the limb-darkening effect of the source with the linear limb darkening law in the modeling. As described in Section~\ref{sec:preamble}, we infer the effective temperature of \(\sim 5000\mathrm{K}\) and \(\log g \sim 4.5\). Assuming the initial turbulence velocity \(\sim 1 \mathrm{km/s}\) and solar metallicity, we have the \(u_{I} = 0.562\) \citep{Claret2011}. Considering the limb-darkening effect, we then investigate the local posterior around these best solutions. 

For this event, the initial grids are sampled based on a reparameterization method similar to \citet{dong2009}, in order to make the grid search more efficient. The initial guesses include 40 values for caustic width $\log{w}$ uniformly spaced over $[-3.5, -2.0]$, 60 values for $\log{q}$ over $[-5.0, 2.0]$ and 10 values for \( \alpha \) over \( [0,2\pi]\). The horizontal caustic width \(\log{w} \) is given by the \citet{chang&refsdal1984} lens approximation with different shear \(\gamma\) definitions and limiting regions \citep{chung2005,an2005,Zhangkeming2023}. Based on these works, a general approximation formula takes the form,
\begin{equation}
w = \frac{4q}{(s-s^{-1})^2 (1+q)^2} (s< s_c),\quad \quad w = \frac{4q}{(s-s^{-1})^2 (1+q)} \quad (s > s_w)
\end{equation}
\begin{equation}
s-s^{-1}- w(s,q) + 2\frac{\gamma q^{\frac{1}{2}}}{{\sqrt{1-\gamma}}} = 0; \qquad (s_c<s < s_w);  \qquad \gamma = {\left( \frac{\sqrt{w^2+4}+w}{2} \right)}^{-2},
\end{equation}
where \(s_c\) and \(s_w\) denote the boundary transition point between the close, wide, and resonant caustics topology \citep{Dominik1999} and follows
\begin{equation}
    s_c = \left[\frac{\left(1-s^4_{c}\right)^3\left( 1+q \right)^2}{27q}\right]^{1/8}; \qquad s_w = \frac{\left(1+q^{1/3} \right)^{3/2}}{\sqrt{1+q}}.
\end{equation}

\begin{figure}
   \centering
   \includegraphics[width=\linewidth]{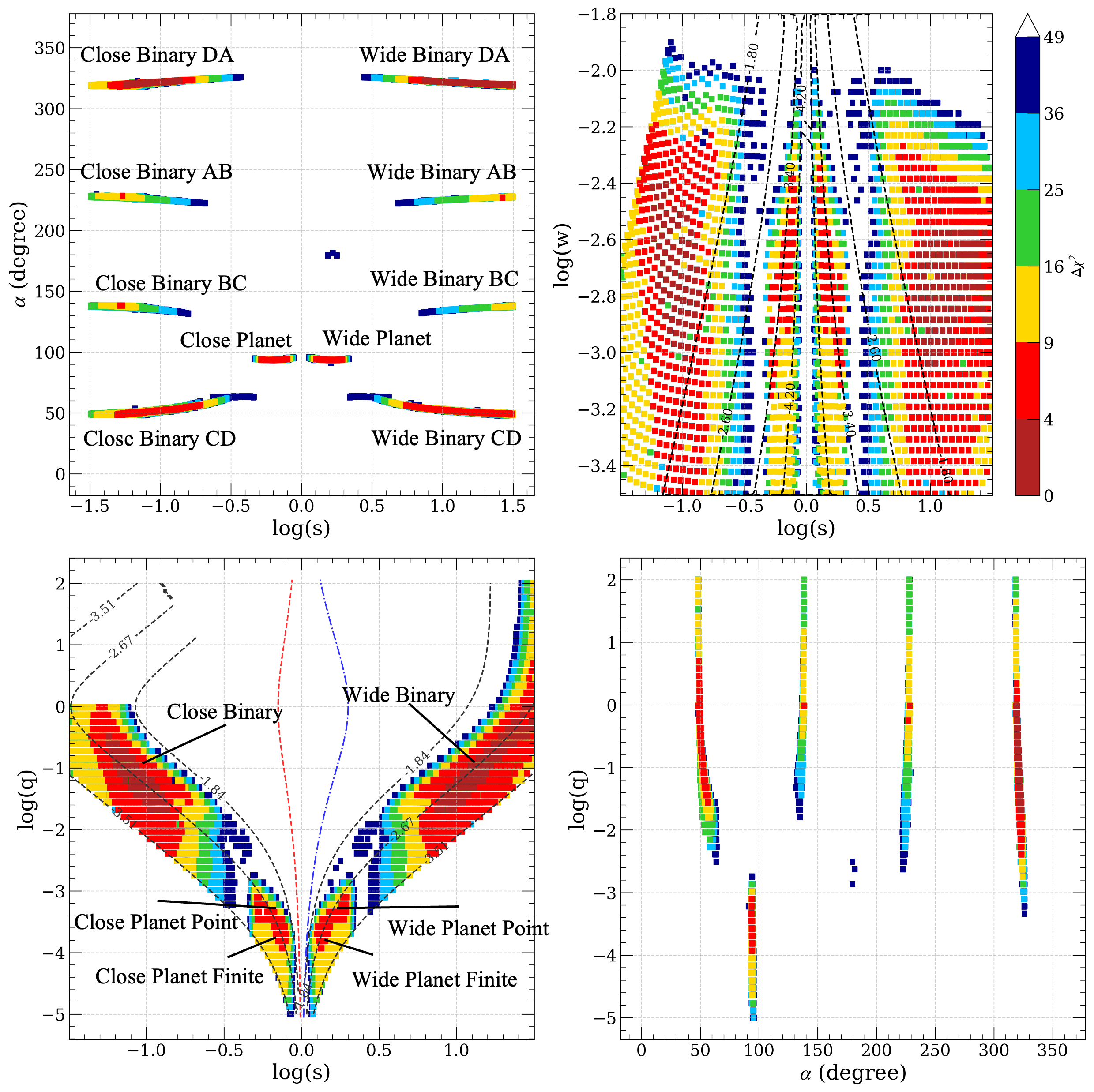}
   \caption{
   The \(\chi^2\) distribution projected onto the four planes of \((\log s, \log q, \alpha, \log w)\) space from the grid search result of \eventa. The grid points with \(\Delta \chi^2 > 49 \) are left blank. The different degenerate solutions are marked with their names. The gray dashed lines in the \( (\log s, \log w)\) and \( (\log s, \log q)\) panels represent the envelopes with constant value of \(\log q\) and \(\log w\), respectively. The red and blue dashed lines mark the boundaries between resonant and close/wide caustics geometries \citep{Dominik1999}. In total, ten local mimina are identified. 
   }
   \label{fig:1314_gridsearch}
\end{figure}

\begin{figure}
    \centering
    \includegraphics[width=\linewidth]{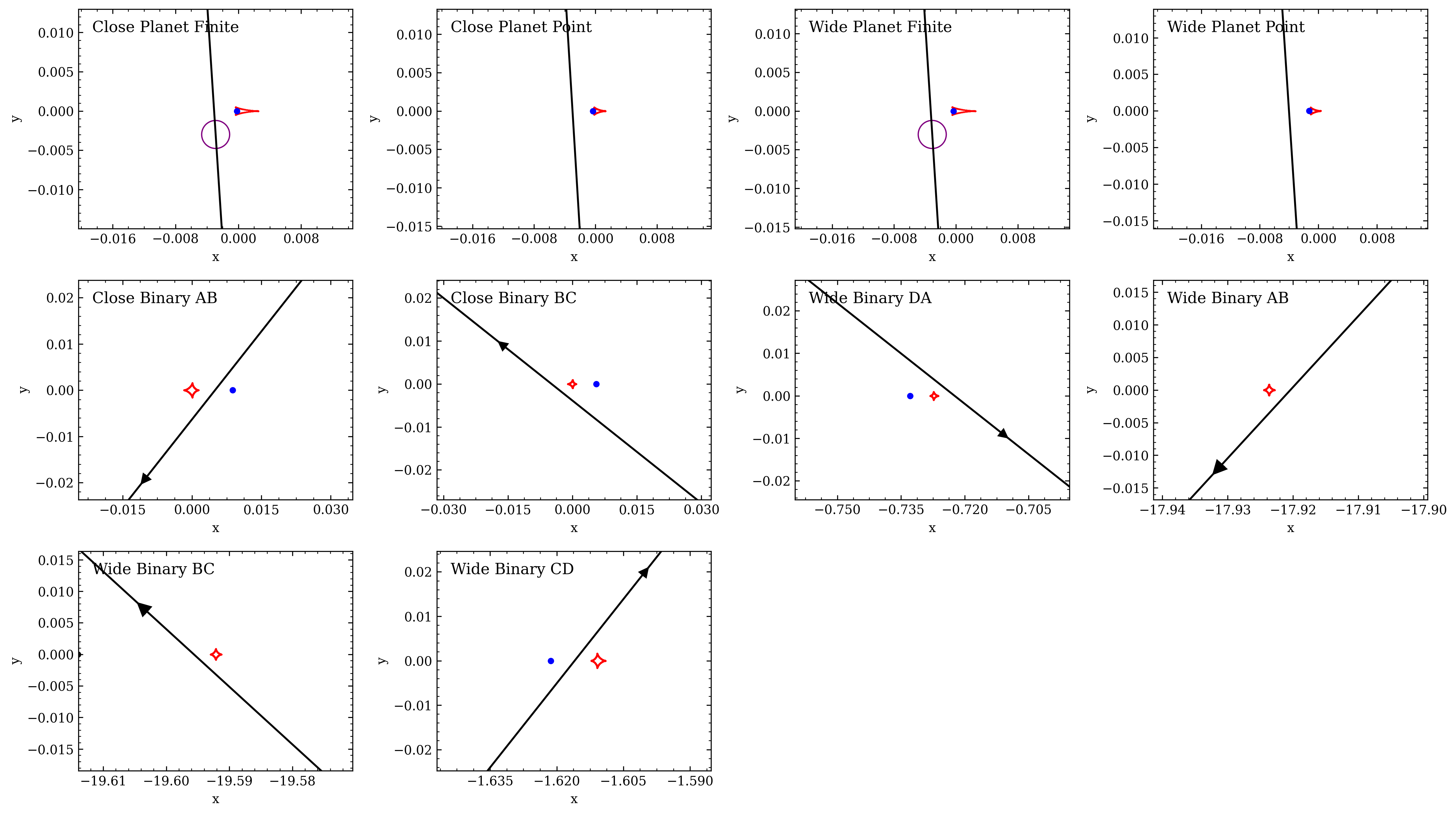}
    \caption{Trajectory configuration of 10 solutions for \eventa. The red curves denote the caustic geometries, the blue dots represent the position of the primary lens, and the purple circles indicate the source radii. The black lines show the source-lens relative trajectory with the arrow indicating the source motion direction. The coordinate origin is set at the magnification center. The 2L1S lensing parameters are shown in Table \ref{tab:1314_2L1S_parameters}.}
    \label{fig:1314_traj}
\end{figure}

The grid-search results are shown in Figure~\ref{fig:1314_gridsearch}. Similar to the three events affected by the ``Planet/Binary'' degeneracy analyzed by \citet{Zhang2025}, \eventa\ has six ``Binary'' solutions, none of which show measurable finite-source effects, as well as four planetary solutions ($\log q < -2$), among which two have measurable finite-source effects, and two do not. This situation is similar to that of KMT-2022-BLG-0954 analyzed by \citet{Zhang2025} and OGLE-2011-BLG-0950 analyzed by Zhang et al.\ (2026). Following the definitions of \citet{Zhang2025}, we label these solutions as ``Close Binary AB'', ``Close Binary BC'', ``Wide Binary AB'',`` Wide Binary BC'', ``Wide Binary CD'', and ``Wide Binary DA'' for the binary solutions, and ``Close Planet Point'', ``Wide Planet Point'', ``Close Planet Finite'', and ``Wide Planet Finite'' for the planetary solutions. Here, ``Close'' denotes solutions with $s<1$, while ``Wide'' denotes solutions with $s>1$. The corresponding caustic structures and source trajectories are shown in Figure~\ref{fig:1314_traj}.
We note that this event represents the third case in which ``Planet Point'' solutions have been identified within the ``Planet/Binary'' degeneracy, following its first identification by \cite{Zhang2025} for the event KMT-2022-BLG-0954. \cite{OB110950_Zhang} subsequently noted the existence of these previously missed solutions in one of the earliest ``Planet/Binary'' events \citep{OB110950}, thereby maintaining the viability of the planetary interpretation when combined with Keck adaptive optics imaging \citep{OB110950_AO}. The discovery of the ``Planet Point'' solutions in the present event demonstrates that this class of solutions is a common feature of the ``Planet/Binary'' degeneracy.

The final best-fit parameters and their uncertainties are listed in Table~\ref{tab:1314_2L1S_parameters}. The modeling indicates that the ``Close Binary BC'' solution provides the best fit to the data, while the other nine solutions are disfavored by $\Delta\chi^2$ values ranging from 0.2 to 14.2. Given the relatively small differences in goodness of fit, we retain all 2L1S solutions and investigate their corresponding source and lens properties in the subsequent section.

We also examine the microlensing parallax effect \citep{Gould1992,Gould2000} by introducing two additional parameters, $(\pi_{\rm E,N}, \pi_{\rm E,E})$, which represent the north and east components of the microlensing parallax vector. However, owing to the faintness of the source ($I_{\rm S} > 22$), the improvement in $\chi^2$ is $<1$, and the $1\sigma$ uncertainties in the parallax components exceed 0.5 in all directions. Consequently, no meaningful constraints on the microlensing parallax can be obtained for this event.

For the 1L2S model, the MCMC-derived parameters are listed in Table~\ref{tab:1L2S_parameters}. The best-fit 1L2S solution is disfavored relative to the best-fit 2L1S model by $\Delta\chi^2 \sim 10$. Finite-source effects are detected for both sources. Based on the source-property analysis presented in the subsequent section, the inferred lens-source relative proper motion is $\mu_{\rm rel} = \theta_{\rm E}/t_{\rm E} \sim 0.5~\mathrm{mas\,yr^{-1}}$. Using Equation~(9) of \citet{2018_subprime}, which is derived from the observed distribution of $\mu_{\rm rel}$ for planetary microlensing events \citep{MASADA}, the probability of obtaining $\mu_{\rm rel} \leq 0.5~\mathrm{mas\,yr^{-1}}$ is only $5 \times 10^{-4}$, corresponding to an effective $\Delta\chi^2 \sim 15$. When this prior information is taken into account, the 1L2S model is disfavored by a total $\Delta\chi^2 \sim 25$. Besides the statistical disfavor, the 1L2S geometry is physically implausible. The projected separation between the two sources ($\Delta u \sim 0.0057$) is only $\sim 1.1$ times the sum of their radii ($\rho_1 + \rho_2 \sim 0.0052$).  For typical stellar masses, this implies a contact or near-contact binary configuration, barring an improbable projection effect. Therefore, we conclude that the 1L2S interpretation is strongly disfavored and physically unrealistic for this event and thus reject the 1L2S model. 

In conclusion, for \eventa, the planetary and stellar-binary interpretations for the 2L1S model both remain viable.

\begin{table}
\begin{center}
\caption[]{2L1S parameters for \eventa}
\label{tab:1314_2L1S_parameters}
\begin{tabular}{lccccccccc}
\hline\noalign{\smallskip}
Model & $t_{0}$ (HJD') & $u_{0}$ & $t_{\rm E}$ (day) & $\alpha$ (degree) & $\log{\rho}$ & $\log{q}$ & $\log{s}$ & $I_{\rm S, KMTA}$ & $\chi^2/\rm{dof}$ \\
\hline\noalign{\smallskip}

Close Planet Finite        
& 10841.212 & 0.00315 & 74.004 & 94.124 & -2.739 & -3.650 & -0.123 & 23.52  & 1022.7/1041 \\
                           
& $\pm$0.002 & $\pm$0.00061 & $\pm$14.421 & $\pm$0.630 & $\pm$0.084 & $\pm$0.118 & $\pm$0.027 & $\pm$ 0.25 &  \\

Close Planet Point         
& 10841.214 & 0.00310 & 71.424 & 93.494 & $<$ -2.665& -3.238 & -0.255 & 23.61 & 1025.3/1041 \\
                           
& $\pm$0.002 & $\pm$0.00053 & $\pm$13.211 & $\pm$0.570 & --- & $\pm$0.097 & $\pm$0.024 & $\pm$ 0.21 &  \\

Wide Planet Finite         
& 10841.213 & 0.00311 & 74.481 & 93.626 & -2.742 & -3.621 &  0.126 & 23.50& 1022.8/1041 \\
                           
& $\pm$0.002 & $\pm$0.00066 & $\pm$15.262 & $\pm$0.658 & $\pm$0.093 & $\pm$0.117 & $\pm$0.028 & $\pm$ 0.22 &  \\

Wide Planet Point          
& 10841.214 & 0.00310 & 70.493 & 93.759 & $<$-2.668& -3.177 &  0.279 & 23.45 & 1025.2/1041 \\
                           
& $\pm$0.001 & $\pm$0.00053 & $\pm$13.066 & $\pm$0.582 & --- & $\pm$0.096 & $\pm$0.023 & $\pm$ 0.18 &  \\

Close Binary AB            
& 10841.207 & 0.004114 & 59.259 & 231.592 & $<$ -2.551 & 0.950 & -1.019 & 23.27 & 1025.7/1041 \\
                           
& $\pm$0.003 & $\pm$0.000769 & $\pm$12.536 & $\pm$1.344 & ---  & $\pm$0.375 & $\pm$0.103 & $\pm$ 0.16 &  \\ %

Close Binary BC          
& 10841.210 & 0.003338 & 66.779 & 141.348 &  $< $-2.595& 1.148 & -1.041 & 23.49 & 1019.5/1041 \\
                           
& $\pm$0.002 & $\pm$0.000570 & $\pm$13.836 & $\pm$1.105 & --- & $\pm$0.296 & $\pm$0.111 & $\pm $ 0.18  &  \\ %

Wide Binary DA            
& 10841.209 & 0.003174 & 70.599 & 321.384 & $<$ -2.756 & -0.977 & 1.134 & 23.45 & 1019.7/1041 \\
                           
& $\pm$0.002 & $\pm$0.000660 & $\pm$19.375 & $\pm$1.037 & --- & $\pm$0.367 & $\pm$0.141 & $\pm$0.22 &  \\ %

Wide Binary AB          
& 10841.217 & 0.002861 & 81.589 & 227.095 &  $<$-2.724 & -0.094 & 1.417 & 23.28 & 1030.6/1041 \\
                           
& $\pm$0.002 & $\pm$0.000503 & $\pm$13.418 & $\pm$0.643 & --- & $\pm$0.240 & $\pm$0.064 & $\pm$ 0.18 &  \\ %

Wide Binary BC          
& 10841.215 & 0.002726 & 87.575 & 137.428 & $<$ -2.617 & 0.168 & 1.440 & 23.24 & 1033.7/1041 \\
                           
& $\pm$0.001 & $\pm$0.000359 & $\pm$9.735 & $\pm$0.521 & --- & $\pm$0.190 & $\pm$0.033 & $\pm$ 0.17 &  \\ %

Wide Binary CD          
& 10841.209 & 0.003732 & 64.885 & 50.832 & $<$ -2.613 & -0.633 & 1.191 & 23.16 & 1025.2/1041 \\
                           
& $\pm$0.002 & $\pm$0.000630 & $\pm$12.383 & $\pm$1.038 & ---  & $\pm$0.317 & $\pm$0.128 & $\pm$ 0.14 &  \\

\hline
\end{tabular}
\end{center}
\tablecomments{\textwidth}{HJD' = HJD - 2450000. The parameters are reported with $1 \sigma$ uncertainties for constrained values and $3 \sigma$ limit for unconstrained $\rho$. The coordinate origin is set at the magnification center.}
\end{table}

\begin{table}[]
\centering
\caption{1L2S parameters for the Two Events}
\label{tab:1L2S_parameters}  
\begin{tabular}{l cc}
\hline\noalign{\smallskip}
Parameter & \eventa & \eventb \\
\hline
$\chi^2 / {\rm dof}$           
& 1029.8/1041                     
& 2032.3/1791 \\
\hline
$t_{0,1}$ (day)   
& $10841.107 \pm 0.004$   
& $10845.177 \pm 0.00019$ \\
$t_{0,2}$ (day)   
& $10841.358 \pm 0.004$   
& $10845.024 \pm 0.005$ \\
$u_{0,1}$          
& $0.00276 \pm 0.00068$   
& $0.00145 \pm 0.00009$ \\
$u_{0,2}$          
& $-0.00236 \pm 0.00070$   
& $-0.01460 \pm 0.00136$ \\
$t_{\rm E}$ (day)  
& $96.34 \pm 18.86$       
& $17.52 \pm 1.34$ \\
$\log\rho_1$       
& $-2.553 \pm 0.108$      
& ... \\
$\log\rho_2$       
& $-2.627 \pm 0.148$      
& $<$ -1.80 \\
$q_{\rm I}$              
& $0.611 \pm 0.097$       
& $2.808 \pm 0.100$ \\
\hline
\end{tabular}

\end{table}

\subsection{KMT-2025-BLG-1392}

\begin{figure}
   \centering
   \includegraphics[width=13.0cm, angle=0]{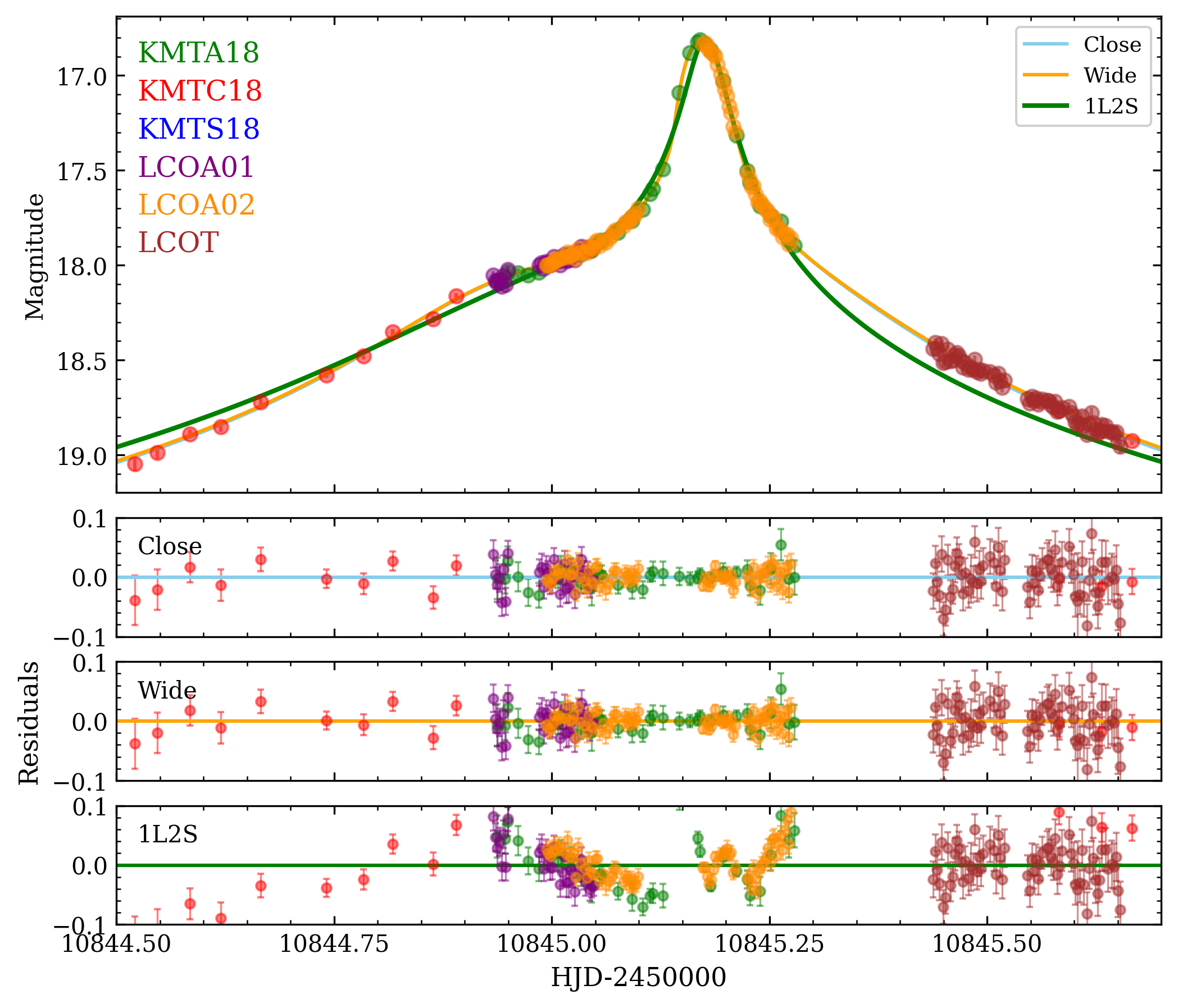}
   \caption{Light curve and the best-fit model of the microlensing event KMT-2025-BLG-1392. The bottom panel shows the residual distribution for the 2L1S close/wide solutions and the 1L2S solution. Note that the multi-site data in the top panel are aligned to the KMTA scale using the source and blend fluxes ($f_s, f_b$) of the``Close`` model. Because the 1L2S model has a significantly different $\chi^2$ and requires different flux parameters, its model curve shows an apparent visual offset from the aligned data. The residuals in the bottom panels, however, are computed independently using each model's respective best-fit flux parameters.}
   \label{fig:1392}
\end{figure}

\begin{figure}
    \centering
    \includegraphics[width=\linewidth]{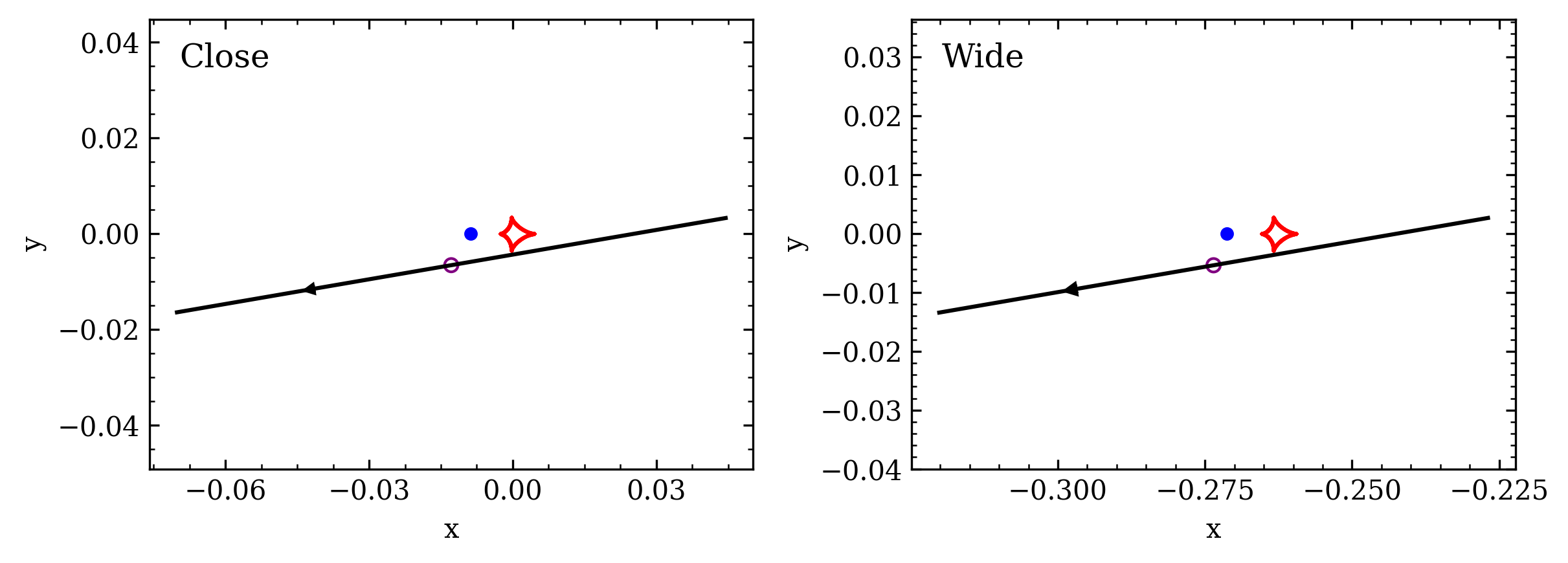}
    \caption{Trajectory configuration of the event KMT-2025-BLG-1392. }
    \label{fig:1392_traj}
\end{figure}

As shown in Figure~\ref{fig:1392}, the light curve of \eventb\ exhibits a $\sim$3-hour bump following the PSPL peak. The anomaly is mainly densely covered by LCOA01, LCOA02, and KMTA data. The adopted limb-darkening coefficient is $u_I = 0.582$ with an effective temperature of \(\sim 4800\mathrm{K}\). The parameter search for \eventb\ begins with a $(40 \times 40 \times 10)$ grid spanning the ranges $\log s \in [-1, 0.7]$, $\log q \in [-5, 0]$, and $\alpha \in [0, 2\pi]$. From the grid-search results, we identify two local minima corresponding to the well-known ``Close/Wide'' degeneracy \citep{Griest1998, Dominik1999, an2005}. The MCMC results with all parameters free are listed in Table~\ref{tab:1392}. The ``Wide'' solution provides the best fit to the data, while the ``Close'' solution is disfavored by only $\Delta\chi^2 = 0.2$. 
Figure~\ref{fig:1392_traj} shows the caustic geometry and the source trajectory, from which it is clear that the anomaly is produced by a source crossing the central caustic.

As shown in Table~\ref{tab:1L2S_parameters} and Figure~\ref{fig:1392}, the 1L2S model fails to reproduce the anomaly and is disfavored relative to the 2L1S model by $\Delta\chi^2 = 246$. We therefore reject the 1L2S interpretation for this event.

With $q \sim 0.05$, the primary lens is inferred to host a companion near the planet/brown-dwarf boundary.

\begin{table}
\begin{center}
\caption[]{2L1S Parameters for \eventb}
\label{tab:1392_2L1S_parameters}
\begin{tabular}{ccccccc}
  \hline\noalign{\smallskip} \label{tab:1392}
Solution & Close & Wide \\
\hline
$\chi^2/ {\rm dof}$     & 1786.2/1791    & 1786.0/1791 \\
\hline
$t_0$(day)          & 10845.145 $\pm$ 0.001 & 10845.146 $\pm$ 0.001 \\
$u_0$               & 0.00424 $\pm$ 0.00052 & 0.00422 $\pm$ 0.00050 \\
$t_{\rm{E}}$(day)   & 25.906 $\pm$ 3.204    & 25.953 $\pm$ 3.477 \\
$\alpha$ (degree)   & 189.866 $\pm$ 0.321   & 189.791 $\pm$ 0.318 \\
$\log{\rho}$        & -2.860 $\pm$ 0.052    & -2.860 $\pm$ 0.055 \\
$q$                 & 0.0434 $\pm$ 0.0078   & 0.0519 $\pm$ 0.0103 \\
$\log{s}$           & -0.691 $\pm$ 0.019    & 0.733 $\pm$ 0.022 \\
$I_{\rm S,KMTC}$ & 22.33 $\pm$ 0.14 & 22.32 $\pm$ 0.13 \\
  \noalign{\smallskip}\hline
\end{tabular}

\end{center}
\end{table}

\section{Source and Lens Properties}
\label{sec:source_lens_properties}
\subsection{Color-Magnitude Diagram}\label{sec:source properties}
\begin{figure}
   \centering
    \includegraphics[width=\linewidth]{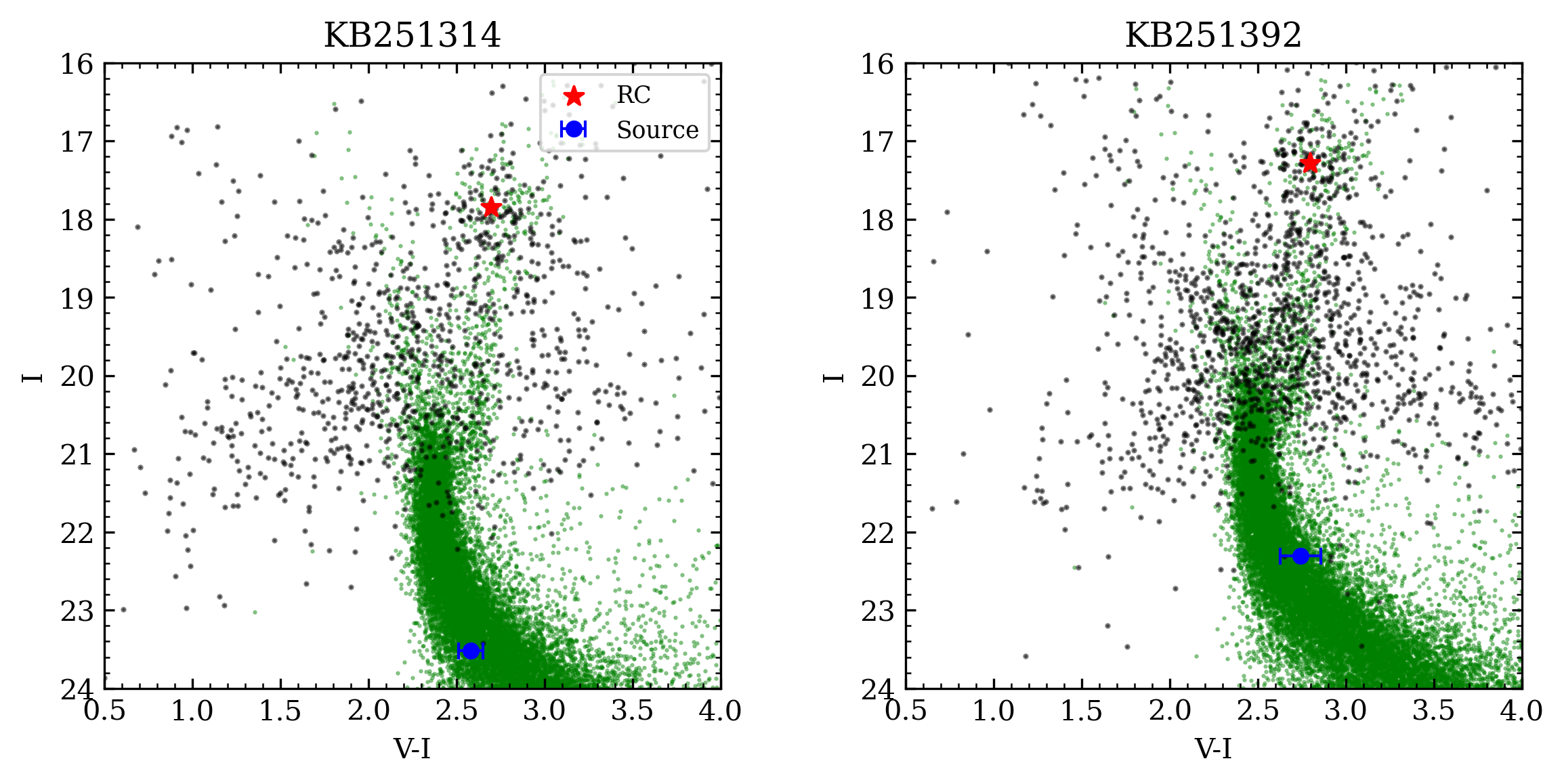}
   \caption{Color magnitude diagram for the two events, constructed from the KMT field stars $120''$ square centered on the event position (KMTA for \eventa\ and KMTC data for \eventb). The red star marker represents the centroid of the red clump, and the blue dot shows the inferred position of the source star from the best-fit 2L1S model. The green dots represent the HST CMD of \citet{holtzman1998} whose red-clump centroid has been matched to that of KMT using $(V - I, I)_{\rm cl, HST} = (1.62, 15.15)$ \citep{MB07192}.
   }
   \label{Fig: CMD}
   \end{figure}

To measure the angular radius of the source star, $\theta_*$, thus to obtain the angular Einstein radius through $\thetae = \theta_*/\rho$, we locate the source star on a color-magnitude diagram (CMD). For both events, the CMD is constructed using the star from the KMT field stars located $120''$ square centered on the event position (KMTA for \eventa\ and KMTC data for \eventb). Following the method in \cite{nataf2013}, we first determine the centroid of the red clump (RC), $(V-I,I)_{(\mathrm{RC},0)}$. From \citet{bensby2013} and Table~1 of \citet{nataf2013}, we obtain the de-reddened color and magnitude of the RC, $(V-I, I)_{\rm RC,0}$. The source $I$-band apparent magnitude is derived from the light-curve modeling. The source color, $(V-I)_{\rm S}$, is calculated based on the linear regression between the $I$-band and $V$-band light curve data. Then, the source de-reddened color and magnitude can be derived by 
\begin{equation}
    (V-I, I)_{\rm S,0} = (V-I, I)_{\rm S} - [(V-I, I)_{\rm RC} - (V-I, I)_{\rm RC,0}]. 
\end{equation}

We apply the empirical color–surface brightness relation of \cite{adams2018} to obtain the source angular size \(\theta_*\). The derived parameters from the above processes, and the inferred $\thetae$ and $\mu_{\rm rel}$ for different solutions are listed in Table~\ref{tab:1314_2L1S_source_property} and Table~\ref{tab:1392_2L1S_source_property}. We also estimate the effective temperature \(T_{\rm{eff}}\) of the source \citep{Houdashelt2000} to determine the limb-darkening coefficients used in Sec~\ref{sec: light_curve_analysis}.

\begin{table}
\begin{center}
\caption{2L1S source properties and derived $\theta_{\rm E}$ and $\mu_{\rm rel}$ for \eventa.}
\label{tab:1314_2L1S_source_property}
\begin{tabular}{lccccc}
\hline\noalign{\smallskip}

Model & $(V - I,I)_{\rm S,KMTA}$ & $(V - I,I)_{\rm S,0}$ & $\theta_* (\mu\text{as})$ & $\theta_{\rm E} (\text{mas})$ & $\mu_{\rm rel} (\text{mas yr}^{-1})$ \\
\hline\noalign{\smallskip}

Close Planet Finite 
& $2.58 \pm 0.07$ & $0.95 \pm 0.08$ & $0.36 \pm 0.05$ & $0.198 \pm 0.048$ & $0.98 \pm 0.30$ \\
& $23.52 \pm 0.25$ & $20.26 \pm 0.27$ & & & \\
\noalign{\smallskip}

Close Planet Point 
& $2.58 \pm 0.07$ & $0.95 \pm 0.08$ & $0.35 \pm 0.04$ & $>0.161$ & $>0.82 $ \\
& $23.61 \pm 0.21$ & $20.35 \pm 0.23$ & & & \\
\noalign{\smallskip}

Wide Planet Finite 
& $2.58 \pm 0.07$ & $0.95 \pm 0.08$ & $0.37 \pm 0.05$ & $0.202 \pm 0.051$ & $0.99 \pm 0.32$ \\
& $23.50 \pm 0.22$ & $20.24 \pm 0.24$ & & & \\
\noalign{\smallskip}

Wide Planet Point 
& $2.58 \pm 0.07$ & $0.95 \pm 0.08$ & $0.37 \pm 0.04$ & $>0.174$ & $>0.90 $ \\
& $23.45 \pm 0.18$ & $20.19 \pm 0.21$ & & & \\
\noalign{\smallskip}

Close Binary AB 
& $2.58 \pm 0.07$ & $0.95 \pm 0.08$ & $0.41 \pm 0.05$ & $>0.144$ & $>0.89 $ \\
& $23.27 \pm 0.16$ & $20.01 \pm 0.19$ & & & \\
\noalign{\smallskip}

Close Binary BC 
& $2.58 \pm 0.07$ & $0.95 \pm 0.08$ & $0.37 \pm 0.04$ & $>0.144$ & $>0.79 $ \\
& $23.49 \pm 0.18$ & $20.23 \pm 0.21$ & & & \\
\noalign{\smallskip}
Wide Binary DA 
& $2.58 \pm 0.07$ & $0.95 \pm 0.08$ & $0.37 \pm 0.05$ & $>0.213$ & $>1.10 $ \\
& $23.45 \pm 0.22$ & $20.19 \pm 0.24$ & & & \\
\noalign{\smallskip}

Wide Binary AB 
& $2.58 \pm 0.07$ & $0.95 \pm 0.08$ & $0.40 \pm 0.05$ & $>0.214$ & $>0.96 $ \\
& $23.28 \pm 0.18$ & $20.02 \pm 0.21$ & & & \\
\noalign{\smallskip}

Wide Binary BC 
& $2.58 \pm 0.07$ & $0.95 \pm 0.08$ & $0.41 \pm 0.05$ & $>0.170$ & $>0.71$ \\
& $23.24 \pm 0.17$ & $19.98 \pm 0.20$ & & & \\
\noalign{\smallskip}

Wide Binary CD 
& $2.58 \pm 0.07$ & $0.95 \pm 0.08$ & $0.43 \pm 0.05$ & $>0.175$ & $>0.99 $ \\
& $23.16 \pm 0.14$ & $19.90 \pm 0.17$ & & & \\

\hline
\end{tabular}
\end{center}
\tablecomments{\textwidth}{The position of the red clump centroid is estimated to be $(V - I, I)_{\rm RC, KMTA} = (2.694 \pm 0.025, 17.850 \pm 0.099)$, and its intrinsic position is $(V - I, I)_{\rm RC,0} = (1.06 \pm 0.03, 14.59 \pm 0.04)$ \citep{bensby2013,nataf2013}.}
\end{table}

\begin{table}
\begin{center}
\caption{2L1S source properties and derived $\theta_{\rm E}$ and $\mu_{\rm rel}$ for \eventb.}
\label{tab:1392_2L1S_source_property}
\begin{tabular}{lccccc}
\hline\noalign{\smallskip}
Model & $(V - I,I)_{\rm S,KMTC}$ & $(V - I,I)_{\rm S,0}$ & $\theta_* (\mu\text{as})$ & $\theta_{\rm E} (\text{mas})$ & $\mu_{\rm rel} (\text{mas yr}^{-1})$ \\
\hline\noalign{\smallskip}

Close 
& $2.74 \pm 0.12$ & $1.01 \pm 0.12$ & $0.56 \pm 0.07$ & $0.407 \pm 0.071$ & $5.74 \pm 1.23$ \\
& $22.33 \pm 0.14$ & $19.42 \pm 0.15$ & & & \\
\noalign{\smallskip}

Wide 
& $2.74 \pm 0.12$ & $1.01 \pm 0.12$ & $0.56 \pm 0.07$ & $0.409 \pm 0.073$ & $5.75 \pm 1.28$ \\
& $22.32 \pm 0.13$ & $19.41 \pm 0.14$ & & & \\

\hline
\end{tabular}
\end{center}
\tablecomments{\textwidth}{The position of the red clump centroid is estimated to be $(V - I, I)_{\rm RC, KMTC} = (2.798 \pm 0.017, 17.284 \pm 0.054)$, and its intrinsic position is $(V - I, I)_{\rm RC,0} = (1.06 \pm 0.03, 14.38 \pm 0.04)$.}

\end{table}

\subsection{Bayesian Analysis}
To determine lens properties in the absence of parallax measurements, we use a Bayesian analysis to incorporate a prior from the galactic model. We follow the basic procedure of \citep{zhu2017}. The differential event rate of the galactic model at a certain source distance \(D_S\) is given by
\begin{equation}
\frac{d^3 \Gamma}{d D_L \, d \log M_L \, d \mu_{\rm rel}} 
=4 \pi n_{\rm L,\star} D_L^2 \theta_{\rm E} \mu_{\rm rel}^2 f_\mu(\mu_{\rm rel}) 
\frac{d\xi(M_L)}{d \log M_L},
\end{equation}
where \(n_{\rm L,\star}\) is the local stellar density given the galactic coordinates, \(f_\mu(\mu_{\rm rel})\) is the probability distribution function for the norm of the lens-source relative proper motion \(\boldsymbol{\mu}_{\mathrm{rel}}\), the $M_{\rm{L}}$ and $D_{\rm{L}}$ are the lens mass and lens distance, respectively. For the lens mass distribution, we use the present-day mass function, which combines the initial mass function of \cite{kroupa2001} at the low-mass end and the mass function for stellar remnants. The mass functions for stellar remnants are adopted from \cite{Gentil2019} for white dwarfs, \cite{Kiziltan2013} for neutron stars, and \cite{olejak2020} for black holes.  For the galactic model, we use the prescription in \cite{zhu2017}.

Because all the event solutions yield no parallax constraints, we use the $\theta_{\rm{E}}$ and $\mu_{\rm{rel}}$ measurements to constrain the posterior. The marginal probability distributions for $D_{\rm{L}}$ and $M_{\rm{L}}$, given certain $\theta_{\rm{E}}$ and $\mu_{\rm{rel}}$ measurements, are derived as follows:
\begin{equation}
    \frac{d^3\Gamma}{dD_{\mathrm{L}}d\theta_{\mathrm{E}}d\mu_{\mathrm{rel}}} = \frac{d^3 \Gamma}{d D_{\mathrm{L}} \, d \log M_{\mathrm{L}} \, d \mu_{\mathrm{rel}}} 
    \left| \frac{\partial (D_{\mathrm{L}}, \log M_{\mathrm{L}} , \mu_{\mathrm{rel}} )}{\partial (D_{\mathrm{L}},\theta_{\mathrm{E}},\mu_{\mathrm{rel}})} \right| \propto n_{\mathrm{L},\star} D_{\mathrm{L}}^2 \mu_{\mathrm{rel}}^2 f_\mu(\mu_{\mathrm{rel}}) 
\frac{d\xi(M_{\mathrm{L}})}{d \log M_{\mathrm{L}}},
\end{equation}
and
\begin{equation}
    \frac{d^3\Gamma}{dM_{\mathrm{L}}d\theta_{\mathrm{E}}d\mu_{\mathrm{rel}}} = \frac{d^3 \Gamma}{d D_{\mathrm{L}} \, d \log M_{\mathrm{L}} \, d \mu_{\mathrm{rel}}} 
    \left| \frac{\partial (D_{\mathrm{L}}, \log M_{\mathrm{L}} , \mu_{\mathrm{rel}} )}{\partial (M_{\mathrm{L}},\theta_{\mathrm{E}},\mu_{\mathrm{rel}})} \right| \propto n_{\mathrm{L},\star} D_{\mathrm{L}}^4 \theta_{\mathrm{E}}^2 \mu_{\mathrm{rel}}^2 M_{\mathrm{L}}^{-1} f_\mu(\mu_{\mathrm{rel}}) 
\frac{d\xi(M_{\mathrm{L}})}{d \log M_{\mathrm{L}}}.
\end{equation}
The marginalized probability of distance can be obtained by integrating the posterior samples from the MCMC chain, i.e, for each sample in the MCMC chains, we calculate the \((\theta_{\mathrm{E}},\mu_{\mathrm{rel}})\) based on the parameters \((t_{\mathrm{E}}, \log \rho,\theta_{\mathrm{\star} })\) and subsequently calculate the marginal distribution of lens mass and lens distance. This gives
\begin{equation}
    \frac{d\Gamma}{D_\mathrm{L}}\ \propto \int \frac{d^3\Gamma}{dD_{\mathrm{L}}d\theta_{\mathrm{E}}d\mu_{\mathrm{rel}}} p(\theta_{\mathrm{E}},\mu_{\mathrm{rel}}|\mathrm{Data})d\theta_{\mathrm{E}}d\mu_{\mathrm{rel}} = \sum_{i \sim \mathrm{MCMC~samples}} \frac{d^3\Gamma}{dD_{\mathrm{L}}d\theta_{\mathrm{E,i}}d\mu_{\mathrm{rel,i}}}.
\end{equation}

Although the finite source effects are not rigorously detected for ``Planet Point'' and ``Binary'' solutions of \eventa, we still use the posterior samples from the MCMC chain with the cut of \(\log \rho \in [-4,-2]\) to avoid unreasonable \((\theta_{\mathrm{E}},\mu_{\mathrm{rel}})\) combinations.
We also consider the variation in $D_{\rm{S}}$ by integration the probability distribution function over all possible values from $D_{\rm{min}} = 6$ kpc to  $D_{\rm{max}} = 10$ kpc with the probability weight $D_{\rm{s}}^{2-\gamma}(\gamma=2.85)$ according to the local stellar density \citep{zhu2017}.

\subsection{KMT-2025-BLG-1314}

For \eventa, the source is likely a K dwarf in the Galactic bulge. For the two ``Planet Finite'' solutions, the lens-source relative proper motion are both low, with $\mu_{\rm rel} \sim 1~\text{mas yr}^{-1}$, for which the values are similar to the other two cases that show have both ``Planet Finite'' and ``Planet Point'' solutions \citep{Zhang2025,OB110950_Zhang}. The ``Planet Finite'' solutions can be verified by future high-resolution imaging by measuring the lens-source proper motions when the lens and the source are resolved. 

The physical parameters derived from the Bayesian analysis for \eventa\ are summarized in Table~\ref{tab:1314_physical}, including the component lens masses, lens distances, and the projected separations between the lens components. For the ``Planet Finite'' solutions, the secondary lens component has a sub-Saturn mass. For the ``Planet Point'' solutions, the companion mass is comparable to that of Saturn. The planet is located near the snowline for the ``Close'' solutions and well beyond the snowline for the ``Wide'' solutions according to the model \(a_{\rm SL} = 2.7 (M/M_\odot)\) \citep{Kennedy2008}. For the ``Binary'' solutions, the secondary lens component falls in the brown-dwarf or M-dwarf mass range, while the primary lens component is an M- or K-dwarf. The lens system is preferentially located in the Galactic bulge.

\begin{table}
\begin{center}
\caption{Physical Parameters Derived from Bayesian Analysis for \eventa.}
\label{tab:1314_physical}
\begin{tabular}{lcccc}
\hline\noalign{\smallskip}
Model 
& $D_{\rm L}$ (kpc) 
& $M_{1}$ ($M_\odot$) 
& $M_{2}$ 
& $r_\perp$ (AU) \\
\hline\noalign{\smallskip}

Close Planet Finite 
& $8.08^{+0.59}_{-0.76}$ 
& $0.38^{+0.32}_{-0.19}$ 
& $0.088^{+0.078}_{-0.051}\,M_{\rm Jup}$ 
& $1.21^{+0.31}_{-0.32}$ \\
\noalign{\smallskip}

Close Planet Point 
& $7.57^{+0.74}_{-1.69}$ 
& $0.61^{+0.36}_{-0.30}$ 
& $0.37^{+0.23}_{-0.20}\,M_{\rm Jup}$ 
& $2.14^{+1.54}_{-1.60}$ \\
\noalign{\smallskip}

Wide Planet Finite 
& $8.09^{+0.61}_{-0.76}$ 
& $0.39^{+0.31}_{-0.20}$ 
& $0.097^{+0.083}_{-0.055}\,M_{\rm Jup}$ 
& $2.18^{+0.59}_{-0.60}$ \\
\noalign{\smallskip}

Wide Planet Point 
& $7.44^{+0.81}_{-2.19}$ 
& $0.64^{+0.36}_{-0.31}$ 
& $0.45^{+0.27}_{-0.24}\,M_{\rm Jup}$ 
& $8.52^{+6.25}_{-6.67}$ \\
\noalign{\smallskip}

Close Binary AB 
& $7.77^{+0.67}_{-1.04}$ 
& $0.48^{+0.32}_{-0.25}$ 
& $0.054^{+0.055}_{-0.050}\,M_\odot$ 
& $0.41^{+0.42}_{-0.42}$ \\
\noalign{\smallskip}

Close Binary BC 
& $7.67^{+0.74}_{-1.32}$ 
& $0.54^{+0.33}_{-0.26}$ 
& $0.039^{+0.034}_{-0.031}\,M_\odot$ 
& $0.69^{+0.58}_{-0.59}$ \\
\noalign{\smallskip}

Wide Binary DA 
& $6.56^{+1.31}_{-3.69}$ 
& $0.67^{+0.34}_{-0.31}$ 
& $0.071^{+0.065}_{-0.063}\,M_\odot$ 
& $92.0^{+81.3}_{-94.6}$ \\
\noalign{\smallskip}

Wide Binary AB 
& $7.41^{+0.82}_{-2.60}$ 
& $0.36^{+0.22}_{-0.19}$ 
& $0.29^{+0.19}_{-0.16}\,M_\odot$ 
& $197.9^{+177.6}_{-189.5}$ \\
\noalign{\smallskip}

Wide Binary BC 
& $7.60^{+0.79}_{-1.60}$ 
& $0.34^{+0.23}_{-0.19}$ 
& $0.23^{+0.16}_{-0.13}\,M_\odot$ 
& $92.9^{+81.6}_{-83.4}$ \\
\noalign{\smallskip}

Wide Binary CD 
& $7.90^{+0.66}_{-0.95}$ 
& $0.39^{+0.29}_{-0.21}$ 
& $0.091^{+0.086}_{-0.072}\,M_\odot$ 
& $72.1^{+71.6}_{-71.8}$ \\

\hline
\end{tabular}
\end{center}
\end{table}

\subsection{KMT-2025-BLG-1392}
The source of \eventb\ is probably a G-dwarf in the Galactic bulge. The measured finite-source effects yield $\thetae \sim 0.4~{\rm mas}$ and $\mu_{\rm rel} \sim 6~\mathrm{mas~yr^{-1}}$, which are typical values for Galactic microlensing events. The physical parameters are listed in Table~\ref{tab:1392_physical}. Both the ``Close'' and ``Wide'' solutions indicate a similar system configuration: a low-mass brown dwarf or (possibly) a massive gas giant orbiting a K- or M-type host star in the Galactic bulge. For the ``Close'' solution, the companion lies inside the snowline, whereas for the ``Wide'' solution it is located well beyond the snowline.

\begin{table}
\begin{center}
\caption{Physical Parameters Derived from Bayesian Analysis for \eventb.}
\label{tab:1392_physical}
\begin{tabular}{lcccc}
\hline\noalign{\smallskip}
Model 
& $D_{\rm L}$ (kpc) 
& $M_{1}$ ($M_\odot$) 
& $M_{2}$ 
& $r_\perp$ (AU) \\
\hline\noalign{\smallskip}

Close 
& $6.89^{+0.61}_{-0.88}$ 
& $0.68^{+0.31}_{-0.27}$ 
& $30.8^{+15.2}_{-13.2}\,M_{\rm Jup}$ 
& $0.57^{+0.11}_{-0.13}$ \\
\noalign{\smallskip}

Wide 
& $6.82^{+0.63}_{-0.92}$ 
& $0.69^{+0.31}_{-0.27}$ 
& $37.7^{+18.5}_{-16.5}\,M_{\rm Jup}$ 
& $15.1^{+3.1}_{-3.5}$ \\

\hline
\end{tabular}
\end{center}
\end{table}

\section{Discussion} \label{sec:disscussion}
In this work, we present the first application of differentiable modeling (e.g., HMC) to binary microlensing events enabled by \texttt{microlux}. In the modeling, the event \eventa\ exhibits a sub-degeneracy for the ``Planet'' solutions characterized by two distinct local minima in the posterior. This poses a challenge for the traditional MCMC method, which is prone to becoming trapped in a certain mode due to strong autocorrelation. Although HMC is not designed for dealing with multimodal posterior, we show that HMC can outperform traditional MCMC because of its weaker autocorrelation during the sampling. 

\begin{figure}
   \centering
    \includegraphics[width=\linewidth]{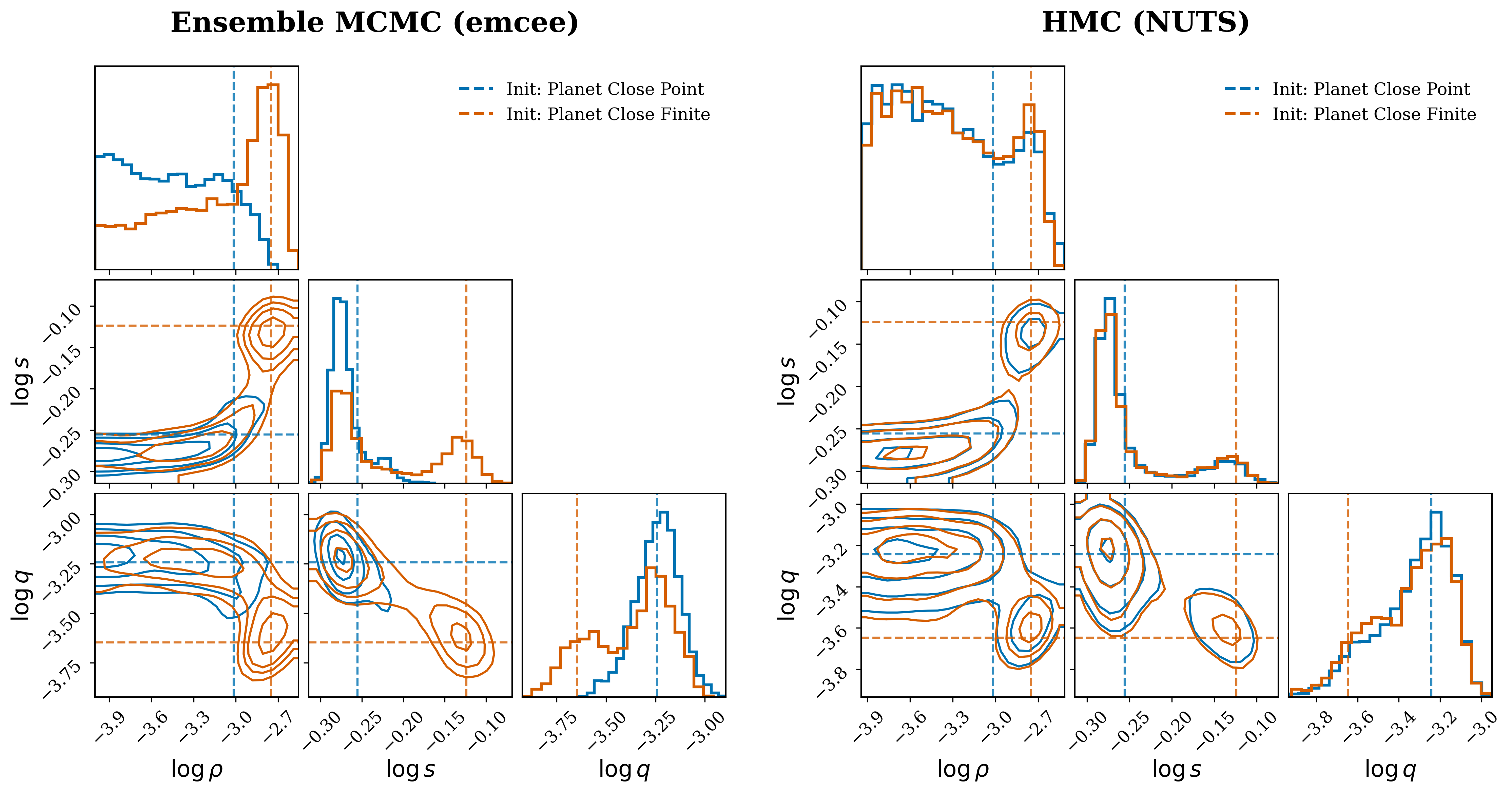}
   \caption{Comparison of posterior sampling between ensemble MCMC sampler (\texttt{emcee}) (left panel) and HMC (NUTS, right panel) for the bimodal posterior of \eventa. The corner plots show the 1D and 2D marginalized posterior distributions for parameters \((\log \rho, \log s, \log q)\). In each panel, two independent chains were initialized at different modes: ``Planet Close Point'' (blue) and ``Planet Close Finite'' (orange), with starting values indicated by dashed lines. 
   Left: The \texttt{emcee} chains fail to converge to a common posterior distribution within the allotted steps. The resulting samples are strongly dependent on their initialization points, indicating poor mixing between modes due to the strong autocorrelation. Right: In contrast, the HMC chains successfully converge to identical posterior distributions regardless of the initialization point. The perfect overlap of contours and histograms demonstrates the robustness and sampling efficiency of HMC in exploring multi-modal landscapes. 
   }
   \label{Fig: HMCvsMCMC}
   \end{figure}

Leveraging the aforementioned Fisher matrix reparameterization procedure in Section~\ref{sec:preamble}, we showcase the divergent behaviors between HMC and ensemble MCMC on sampling the bimodal distribution of \eventa. Figure~\ref{Fig: HMCvsMCMC} shows the comparison between NUTS in HMC (\texttt{Numpyro}) \citep{bingham2019pyro,phan2019composable} and the affine-invariant ensemble sampler of MCMC (\texttt{emcee}) \citep{foreman-mackey2013}. For each sampling method, we initialize the sampler at different modes to test whether the sampler can overcome the potential barrier and reveal the true posterior. The result indicates that the HMC samples with different initial points show a consistent posterior distribution, while the MCMC samples show discrepancy. This inconsistency arises because the ensemble sampler fails to mix effectively between the separated modes. This lack of convergence can be quantitatively diagnosed using the Gelman-Rubin statistic, $\hat{R}$ \citep{Gelman1992, Vehtari2021}, which compares the parameter covariance consistency between and within chains. For the case shown in Figure~\ref{Fig: HMCvsMCMC}, $\hat{R} \sim 1.3 $ for \texttt{emcee} results and $\hat{R} = 1 $ for HMC results. The MCMC chain samples are collected with 40 walkers, 2000 warm-up steps, and 4000 sample steps for each chain. The HMC chain samples are collected with 1 chain, 1000 warm-up steps, 4000 sample steps, and 0.9 target acceptance probability.

The above case demonstrates that HMC outperforms traditional MCMC in sampling more complex posterior distributions, even though HMC itself may not work for all types of multi-modal posteriors. For the more complicated situations, other advanced sampling algorithms are needed,
such as the tempering method, Sequential Monte Carlo, and Nested sampling. 
The implementation of these algorithms will also benefit from a differentiable model such as  \texttt{microlux}.
The new modeling paradigm enabled by such differentiable models may play a crucial role in the upcoming microlensing datasets from next-generation surveys.

\normalem
\begin{acknowledgements}
This work is supported by the National Natural Science Foundation of China (grant Nos.\ 12133005 and 12173021). This research has made use of the KMTNet system operated by the Korea Astronomy and Space Science Institute (KASI) at three host sites of CTIO in Chile, SAAO in South Africa, and SSO in Australia. Data transfer from the host site to KASI was supported by the Korea Research Environment Open NETwork (KREONET). This research was supported by KASI under the R\&D program (project No. 2025-1-830-05) supervised by the Ministry of Science and ICT. H.Y. acknowledges support by the China Postdoctoral Science Foundation (No. 2024M762938). Work by J.C.Y. acknowledges support from N.S.F Grant No. AST-2108414. Work by C.H. was supported by the grants of the National Research Foundation of Korea (2019R1A2C2085965 and 2020R1A4A2002885). Y.S. acknowledges support from BSF Grant No. 2020740. This work is part of the ET space mission, which is funded by China's Space Origins Exploration Program.
The authors acknowledge the Tsinghua Astrophysics High-Performance Computing platform at Tsinghua University for providing computational and data storage resources that have contributed to the research results reported within this paper.

\ 
\end{acknowledgements}

\bibliographystyle{raa}
\bibliography{bibtex}

\end{document}